\documentstyle[twoside,fleqn,espcrc2,graphicx]{article}

\newcommand{\AmS}{{\protect\the\textfont2
  A\kern-.1667em\lower.5ex\hbox{M}\kern-.125emS}}

\title{Direct Detection of WIMP Dark Matter
        \thanks{Review Talk given at the {\em Sixth International Workshop
        on Topics in Astroparticle and Underground Physics}, TAUP 99 (Paris,
        College de France, September 6-10, 1999), to be published in Nucl.
        Phys. B (Proc. Suppl.).}}

\author{Angel Morales\address{Laboratory of Nuclear and High Energy Physics \\
        University of Zaragoza \\
        50009 Zaragoza. Spain}}%

\begin{document}

\begin{abstract}
The status of the recent efforts in the direct search for Weak Interacting
Massive Particle (WIMP) Dark Matter is briefly reviewed and the main
achievements illustrated by the contributions presented to TAUP 99. The
strategies followed in the quest for WIMPs will be first revisited and then
the new results and the future prospects reported.
\end{abstract}

\maketitle

\section{Introduction}

There is substantial evidence, reviewed at length in this Workshop, that
most of the matter of the universe is dark and a compelling motivation
to believe that it consists mainly of non-baryonic objects. From the
cosmological point of view, two big categories of non-baryonic dark
matter have been proposed: cold (WIMPs, axions) and hot (light
neutrinos) dark matter according to whether they were slow or fast
moving at the time of galaxy formation. Without entering into
considerations about how much of each component is needed to fit
better the observations, nor about how large the baryonic component of
the galactic halo could be, we assume that there is enough room for
WIMPs in the halo to try to detect them, either directly or through their
by-products. Discovering this form of dark matter is one of the big
challenges in Cosmology, Astrophysics and Particle Physics.

The indirect detection of WIMPs proceeds currently through two main
experimental lines: either by looking in the space for positrons,
antiprotons, or other antinuclei produced by the WIMPs annihilation in
the halo, or by searching in large underground detectors or underwater
neutrino telescopes for upward-going muons produced by the energetic
neutrinos emerging as final products of the WIMPs annihilation in
celestial bodies (Sun, Earth...)

The direct detection of WIMPs relies in the measurement of the WIMP
elastic scattering off the target nuclei of a  suitable detector. Pervading
the galactic halos, slow moving ($\sim300$ km/s), and heavy ($10 \sim
10^3$ GeV) WIMPs could make a nucleus recoil with a few keV ($\rm T
\sim (1-100)$ keV), at a rate which depends of the type of WIMP and
interaction. Only a fraction of the recoil QT is visible in the detector,
depending on the type of detector and target and on the mechanism of
energy deposition. The so-called Quenching Factor Q is essentially unit
in thermal detectors whereas for the nuclei in conventional detectors
range from about 0.1 to 0.6. Because of the low interaction rate
[typically $<$ 0.1 (0.001) c/kg day for spin independent (dependent)
couplings] and the small energy deposition, the direct search for particle
dark matter through their scattering by nuclear targets requires
ultralow background detectors of a very low energy threshold. Moreover,
the (almost) exponentially decreasing shape of the predicted nuclear
recoil spectrum mimics that of the low energy background registered by
the detector. All these features together make the WIMP detection a
formidable experimental challenge.

Customarily, one compares the predicted event rate with the observed
spectrum. If the former turns out to be larger than the measured one,
the particle under consideration can be ruled out as a dark matter
component. That is expressed as a contour line $\sigma$(m) in the
plane of the WIMP-nucleus elastic scattering cross section versus the
WIMP mass which excludes, for each mass m, those particles with a
cross-section above the contour line $\sigma$(m). The level of
background limits, then, the sensitivity to exclusion.

This mere comparison of the expected signal with the observed
background spectrum is not supposed to detect the tiny imprint left by
the dark matter particle, but only to exclude or constrain it. A
convincing proof of the detection of WIMPs would need to find unique
signatures in the data characteristic of them, like temporal (or other)
asymmetries, which cannot be faked by the background or by
instrumental artifacts. The only distinctive signature investigated up to
now is the predicted annual modulation of the WIMP signal rate.

The detectors used so far (most of whose results have been presented to
this Workshop) are Ge (IGEX, COSME, H/M) and Si (UCSB) diodes, NaI
(ZARAGOZA, DAMA, UKDMC, SACLAY, ELEGANTS), Xe (DAMA, UCLA,
UKDMC) and CaF$_2$ (MILAN, OSAKA, ROMA) scintillators, Al$_2$O$_3$
(CRESST, ROSEBUD) and TeO$_2$ (MIBETA, CUORICINO) bolometers
and Si (CDMS) and Ge (CDMS, EDELWEISS) thermal hybrid detectors,
which also measure the ionization. But new detectors and techniques
are entering the stage. Examples of such new devices, presented at
TAUP are: a liquid-gas Xenon chamber (UCLA); a gas chamber sensitive
to the direction of the nuclear recoil (DRIFT); a device  which uses
superheated droplets (SIMPLE), and a colloid of superconducting
superheated grains (ORPHEUS). There is also some new projects
featuring a large amount of target nuclei, both with ionization Ge
detectors (GENIUS, GEDEON) and cryogenic thermal devices (CUORE).

\section{Strategies for WIMP detection}
The rarity and smallness of the WIMP signal dictate the experimental
strategies for its detection:

Reduce first the background, controlling the radiopurity of the detector,
components, shielding and environment. The best radiopurity has been
obtained in the Ge experiments (IGEX, H/M, COSME). In the case of the
NaI scintillators, the backgrounds are still one or two orders of
magnitude worse than in Ge (ELEGANTS, UKDMC, DAMA, SACLAY).
The next step is to use discrimination mechanisms to distinguish
electron recoils (tracers of the background) from nuclear recoils
(originated by WIMPs or neutrons). Various techniques have been
applied for such purpose: a statistical pulse shape analysis (PSD) based
on the different timing behaviour of both types of pulses (DAMA,
UKDMC, SACLAY); an identification on an event-by-event basis of the
nuclear recoils by measuring at the same time two different
mechanisms of energy deposition, like the ionization and heat
capitalizing the fact that for a given deposited energy (measured as
phonons) the recoiling nucleus ionizes less than the electrons (CDMS,
EDELWEISS).

A promising discriminating technique is that used in the liquid-gas
Xenon detector with ionization plus scintillation presented to this
Workshop (see D. Cline's contribution to these Proceedings). An electric
field prevents recombination, the charge being drifted to create a second
pulse in the addition to the primary pulse. The amplitudes of both
pulses are different for nuclear recoils and gammas allowing their
discrimination.

Another technique is to discriminate gamma background from neutrons
(and so WIMPs) using threshold detectors---like neutron
dosimeters---which are blind to most of the low Linear Energy Transfer
(LET) radiation (e, $\mu$, $\gamma$). A new type of detector
(SIMPLE) which uses superheated droplets which vaporize into bubbles
by the WIMP (or other high LET particles) energy deposition has been
presented to this Workshop (see J. Collar's contribution to these
Proceedings).

The other obvious strategy is to make detectors of very low energy
threshold and high efficiency to see most of the signal spectrum, not
just the tail. That is the case of bolometer experiments (MIBETA,
CRESST, ROSEBUD, CUORICINO, CDMS, EDELWEISS).

Finally, one should search for distinctive signatures of the WIMP, to
prove that you are seeing a WIMP. Suggested identifying labels are: an
annual modulation of the signal rate and energy spectrum (of a few
percent) due to the seasonal June-December variation in the relative
velocity Earth-halo and a forward-backward asymmetry in the direction
of the nuclear recoil due to the Earth motion through the halo. The
annual modulation signature has been already explored. Pioneering
searches for WIMP annual modulation signals were carried out in
Canfranc, Kamioka and Gran Sasso. At TAUP 97, the DAMA experiment
at Gran Sasso, using a set of NaI scintillators reported an annual
modulation effect interpreted (after a second run) as due to a WIMP of
60 GeV of mass and scalar cross-section on protons of
$\sigma_{\rm p} = 7 \times 10^{-6}$ picobarns.

The implementation of these strategies will be illustrated by a selection
of the experiments presented at TAUP 99. The characteristic features
and main results of these experiments are overviewed in the following
paragraphs.

\section{Germanium Experiments with conventional detectors}
The high radiopurity and low background achieved in Germanium
detectors, their fair low energy threshold, their reasonable Quenching
Factor (25\%) and other nuclear merits make Germanium a good option
to search for WIMPs with detectors and techniques fully mastered. The
first detectors applied to WIMP direct searches (as early as in 1987)
were, in fact, Ge diodes, as by-products of $2 \beta$-decay dedicated
experiments. The exclusion plots $\sigma$(m) obtained by former Ge
experiments [PNNL/USC/Zaragoza (TWIN and COSME-1), UCSB,
CALT/NEU/PSI, H/M] are still remarkable and have not been
surpassed till recently.

There are three germanium experiments currently running for WIMP
searches (COSME-2, IGEX and H/M).

COSME-2 (Zaragoza/PNNL/USC) is a small (240 g) natural abundance
germanium detector of low energy threshold (1.8--2 keV) and energy
resolution of 400 eV at 10 keV. It has been underground for more than
ten years and so is rather clean of the cosmogenic induced activity in
the 8--12 keV region. It is currently taking data in Canfranc (at 2450
m.w.e.) for WIMPs and solar axion searches (see I.G. Irastorza's
contribution to these Proceedings). The background is 0.6 (0.3) c/(keV
kg day) averaged between the 2--15 keV (15--30 keV) energy region.

\begin{figure}[t]
\includegraphics[width=7.2cm]{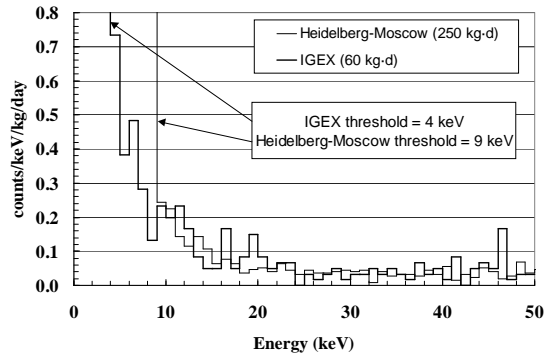}
\caption{IGEX and H/M spectra in the low energy region.}
\label{fig:fig1}
\end{figure}

IGEX is a set of enriched Ge-76 detectors looking for $2 \beta$ decay
(see D. Gonzalez's contribution to these Proceedings) which have been
recently upgraded for WIMP searches with energy thresholds of $\leq$ 4
keV, and energy resolution of 2 keV (at 10 keV). Data from one of these
detectors (RG2, 2.1 kg $\times 30$ d) show backgrounds of 0.1 c/(keV
kg day) in the 10-20 keV region and 0.04 c/(keV kg day) between 20
and 40 keV. It is remarkable that below 10 keV, down to the threshold
of 4 keV the background is $\sim 0.3$ c/(keV kg day) mainly from
noise which is being removed. The spectrum is shown in Fig. \ref{fig:fig1}.
IGEX is also operating two other Ge detectors in Baksan [one
natural---TWIN---and other enriched in Ge-76 (RV)] of about 1 kg each,
and thresholds of 2 and 6 keV respectively. After a subtraction
procedure a background of $\sim 0.1$ c/(keV kg day) between 10 and
30 keV was obtained.

The Heidelberg/Moscow Ge experiment on $2 \beta$ decay in Gran
Sasso is also using data of one of their enriched Ge-76 detectors (2.7
kg, and threshold of 9--10 keV) in a search for WIMPs. The background
is similar to that of IGEX (0.16 c/(keV kg day) from 10-15 keV and 0.05
c/(keV kg day) from 15-30 keV), although its threshold is more than a
factor two higher. The low-energy spectrum is also shown comparatively
to that of IGEX in Fig. \ref{fig:fig1}.

\begin{figure}[htb]
\includegraphics[width=7.2cm]{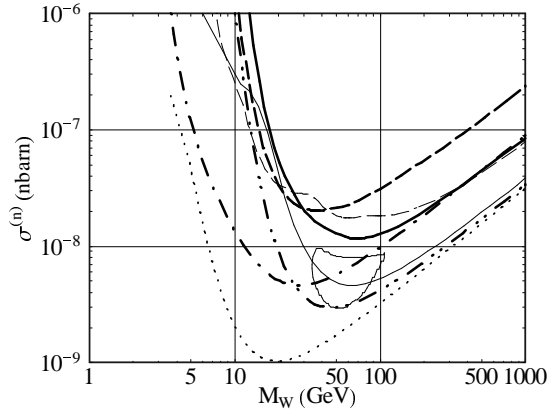}
\caption{Combined exclusion plots obtained from previous
Ge experiments (thin dashed line) compared with those presented to
this Workshop---CDMS (dot-dashed), COSME (thick dashed), IGEX (thick
solid)---and with the DAMA results (thin solid). Prospects for IGEX
(dot-dot-dashed) and CUORICINO (dots) are also shown}
\label{fig:fig2}
\end{figure}

The exclusion plots obtained from the spectra of Germanium
detectors are shown in Fig. \ref{fig:fig2}. The Ge-combined
limit is the contour obtained from the envelope of all of them---including
the last H/M data  and is compared with that derived from the last COSME,
IGEX and CDMS  data presented at this Workshop. Also the most stringent
NaI exclusion plot is shown together with the ($\sigma$, m) region where
a seasonal modulation effect in the recorded rate has been reported by
the DAMA Collaboration and attributed to a WIMP signal. The exclusions
depicted in this paper refer to
spin-independent interactions. The sensitivity of the present detectors
does not yet reach the rates needed to explore spin-dependent
couplings. For comparison among different experiments, the coherent
spin independent WIMP-nucleus cross-section is normalized to that of
WIMP on nucleons. All the Ge and NaI exclusion plots shown have been
recalculated from the original spectra by the author and his
collaborators I.G. Irastorza and S. Scopel, with the same set of
parameters. The values used for the halo model are $\rho = 0.3$ GeV
cm$^{-3}$, $v_{\rm rms} = 270 \rm Kms^{-1}$ and $v_{\rm esc} = 650
\rm Km \rm s^{-1}$.

There exist some projects with Ge detectors in different degree of
development: HDMS (Heidelberg DM Search) is a small (200 g) Ge
detector placed in a well-type large (2 kg) Ge crystal. The goal is to reach
a background of $10^{-2}$ c/(keV kg day). GEDEON (Germanium Diodes
in One Cryostat, Zaragoza / USC / PNNL) will use a single cryostat of
IGEX technology hosting a set of natural Ge crystals of total mass
of 28 kg. The threshold of each small detector is $< 2$ keV and the
background goal---expected from the measured radioimpurities---is
$10^{-2} - 10^{-3}$ c/(keV kg day). A set of three cryostats ($\sim 80$  kg
of Germanium) is the planned final configuration which, embedded in Roman
lead and graphite, will be installed in Canfranc. GENIUS (Germanium
Detectors in Liquid Nitrogen in an Underground Set-up) plans to
operate 40 natural abundance, naked germanium crystals (of 2.5 kg
each) submerged directly in a tank of liquid nitrogen.

\section{Sodium Iodine experiments}
The full isotopic content on A-odd isotopes (Na-23, I-127) makes sodium
iodine detectors sensitive also to spin-dependent WIMP interactions.
The main recent interest in scintillators is due to the fact that
large masses of NaI
crystals for exploring the annual modulation are affordable. There exist
four NaI experiments running (UKDMC, DAMA, SACLAY and
ELEGANTS V) and two in preparation (ANAIS and NAIAD).

The NaI experiments serve to illustrate one of the strategies for
background discrimination mentioned above. The time shape
differences between electron recoils and nuclear recoils pulses in NaI
scintillators can be used to discriminate gamma background from
WIMPs (and neutrons) because of the shorter time constants of nuclear
versus electron recoils. Templates of reference pulses produced by
neutron, gamma (and X, $\beta$...) sources are compared with the data
population pulses (in each energy band) by means of various
parameters [time constants (UKDMC), momenta of the time distribution
(DAMA, SACLAY), integrated time profile differences (SACLAY)]. From
this comparison, the fraction of data which could be due to
nuclear recoils turns out to be only a few percent, depending on
energy, of the measured background [1 to 3 c/(keV kg day) in DAMA
and UKDMC, and of 2 to 10 c/(keV kg day) in SACLAY], with
different degree of success, depending of the experiment and (slightly)
on the method used. Due to the drastic background
reduction, the exclusion plots obtained from the stripped spectra have
surpassed (DAMA, UKDMC) that derived from the (non-manipulated)
spectra of the Ge detectors---whose radiopurity is much better than
that of NaI. Let us briefly mention the main performances of these experiments.

\begin{figure}[htb]
\includegraphics[width=7.2cm]{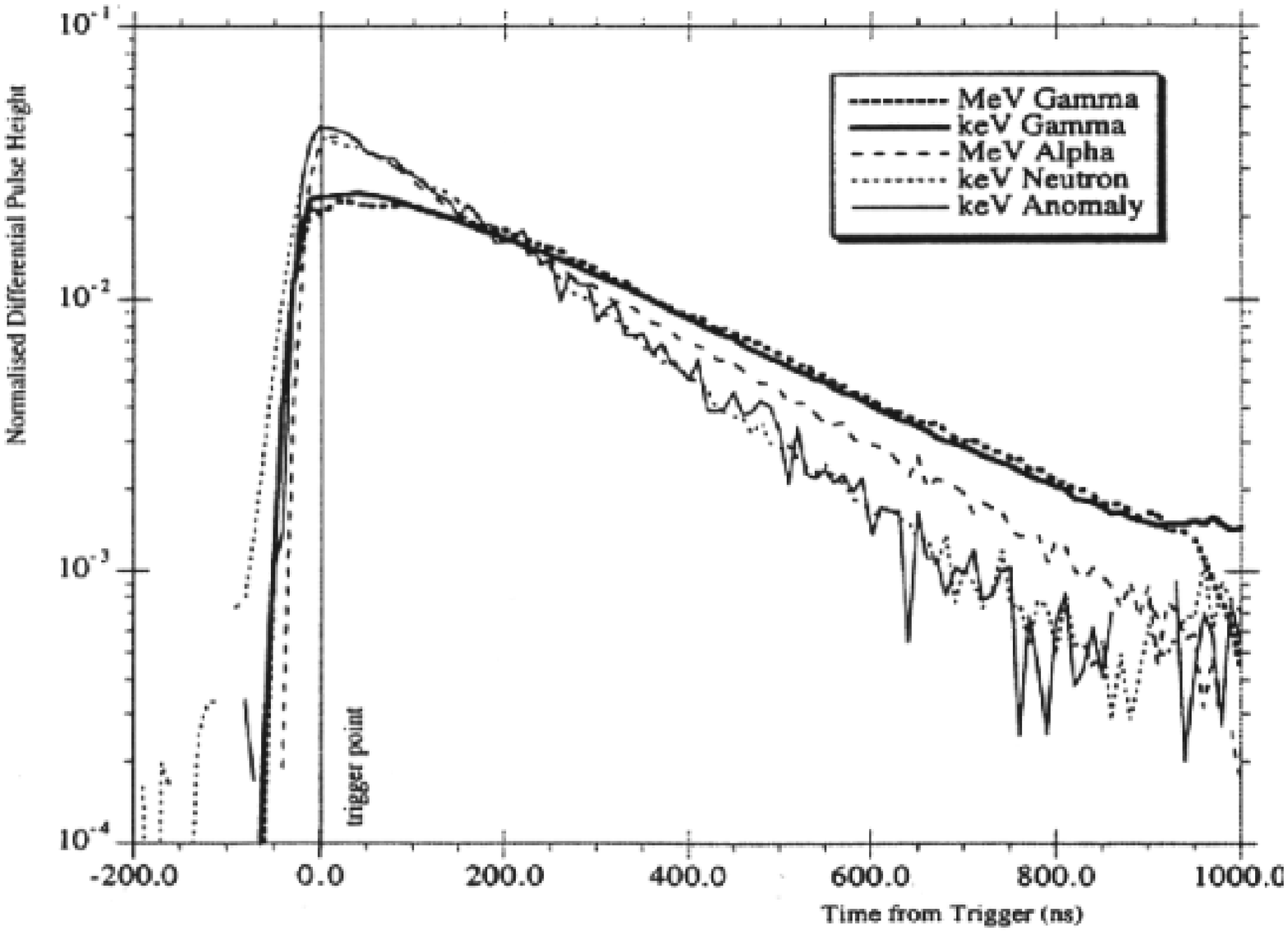}
\caption{Pulse shapes for gammas, alphas, neutrons and the
anomalous events from a 2-kg NaI crystal (UKDMC)}
\label{fig:fig4}
\end{figure}

The United Kingdom Dark Matter Collaboration (UKDMC) uses
radiopure NaI crystals of various masses (2 to 10 kg) in various
shielding conditions (water, lead, copper) in Boulby. Typical thresholds
of 4 keV and background (before PSD) of 2--4 c/(keV kg day) (at about
threshold) have been obtained. Recent results from NaI crystals of
5 and 2 kg show a small population of pulses (Fig. \ref{fig:fig4}) of an
average time constant shorter than that of
gamma events and near to that corresponding to neutron-induced recoils,
which is not due to instrumental artifact. (For recent results, see
I. Liubarsky's contribution to
these Proceedings). Plans of the UKDMC include NAIAD (NaI Advanced
Detector) consisting of 50--100 kg in a set of unencapsulated crystals to
avoid surface problems and improve light collection.

\begin{figure}[htb]
\includegraphics[width=7.2cm]{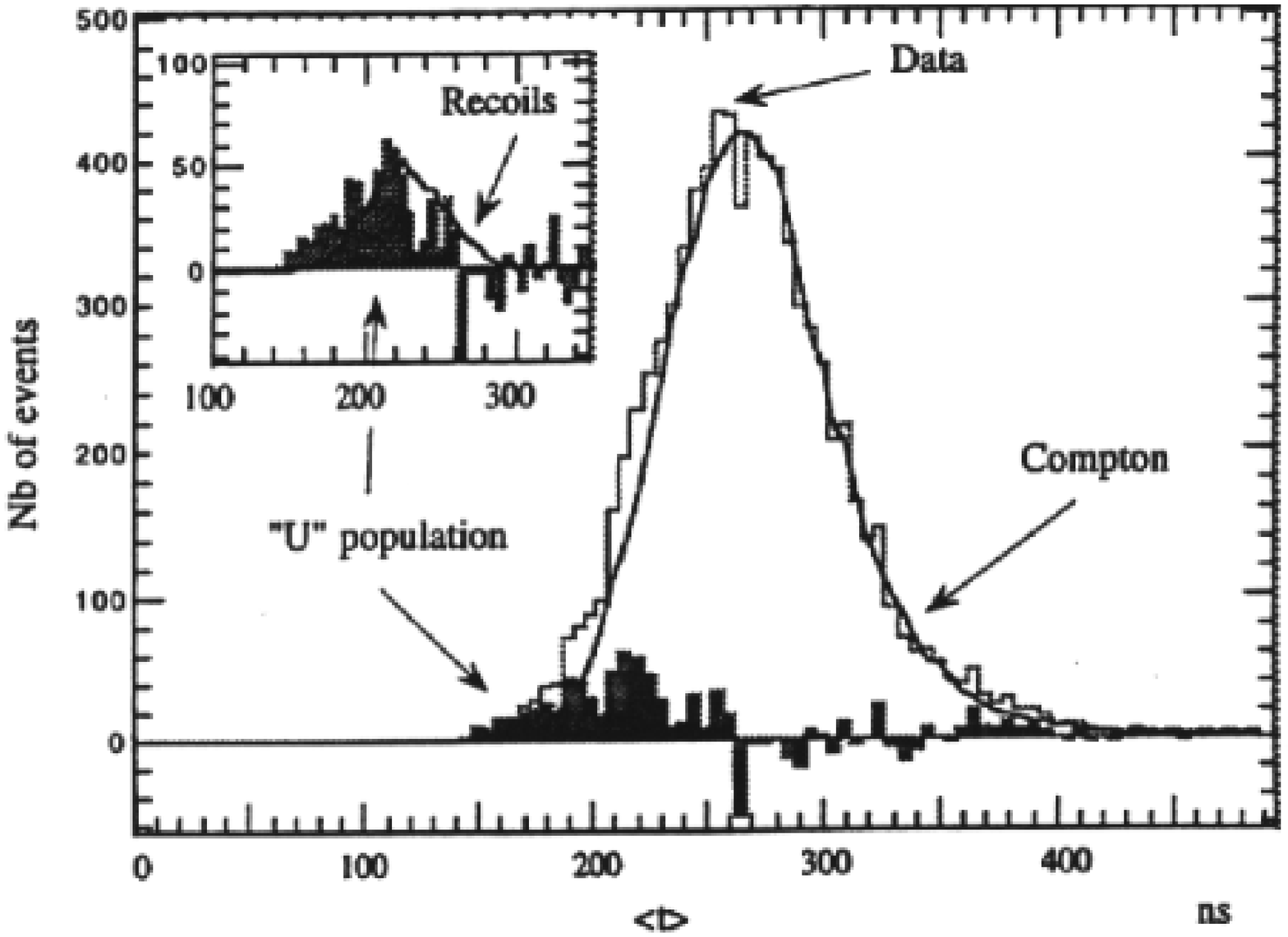}
\caption{$<t>$ distribution for the 10--15 keV bin showing unknown
population ``U'' responsible for shorter decay times in the time profiles
(SACLAY)}
\label{fig:fig5}
\end{figure}

The SACLAY group is carrying out a thorough program of investigation
about the virtues and limitations of the pulse shape analysis as far as
the statistical background discrimination is concerned. They use, at
LSM Frejus, a radiopure 9.7 kg NaI crystal with an energy threshold of
2 keV and backgrounds of (before PSD) 8--10 c/(keV kg day) (at 2--3 keV)
and of $\sim 2$ c/(keV kg day) (and flat) above 5 keV. The high
background at threshold, not well understood, has
spoiled the exclusion plots of this experiment, compared with other NaI
searches. On the other hand, their data---as it happened in that of
UKDMC---cannot be sharply split into Compton plus nuclear recoils
showing a spurious population (Fig. \ref{fig:fig5}), a fact that limits the
sensitivity of the PSA they could perform. Including this peculiar
situation, as systematic effects, their PSA background reduction is only
65\% to 85\%.

The DAMA experiment uses also NaI crystals of 9.7 kg with
energy threshold of $\sim 2$ keV. No spurious population is
found in DAMA which could spoil the PSA separation
background/nuclear recoils. In fact, their background
reduction reaches levels of 85\% (4--6 keV) and 97\% (12--20
keV), providing exclusion plots which have surpassed that of
germanium.

\begin{figure}[htb]
\includegraphics[width=7.2cm]{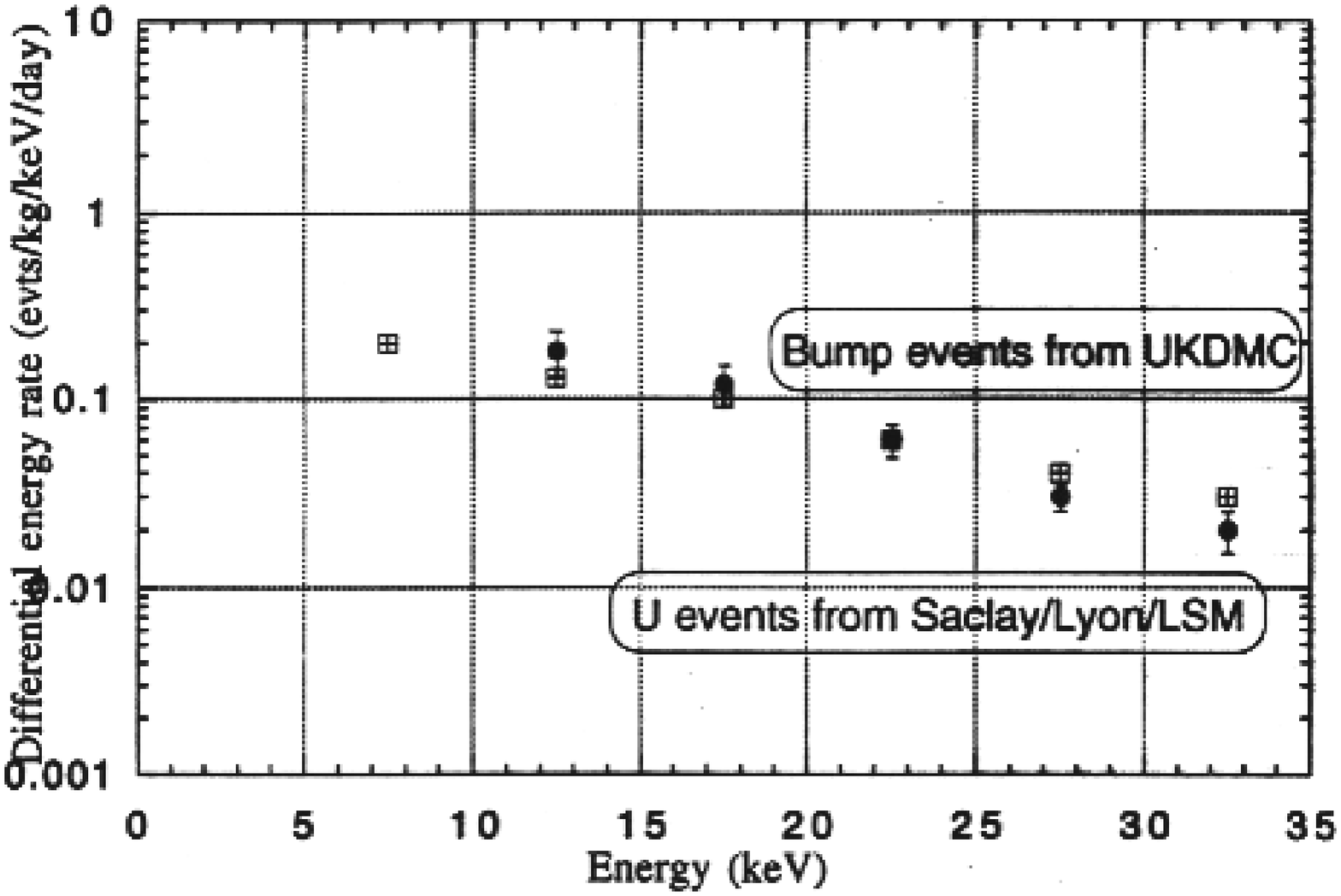}
\caption{Energy spectra of the unknown population
of events found in NaI experiments from UKDMC and SACLAY}
\label{fig:fig6}
\end{figure}

In conclusion, besides the significant reduction of background provided
by this statistical method, a most intriguing result, as stated above, is
that UKDMC and SACLAY, applying these PSD techniques to their data,
have found that they are not compatible with a contribution of only
Compton nor with nuclear recoil events, and suggest the existence of an
unknown population or systematic effects. It is also an intriguing
coincidence that the energy spectrum of that residual populations are
very similar in both experiments (Fig. \ref{fig:fig6}). (See G. Gerbier's
contribution to these Proceedings).

\section{WIMPs Annual Modulation as a distinctive signature}
The Earth rotation around the sun combined with the sun velocity
through the halo makes the velocity of the detector with respect to the
WIMPs halo a yearly periodic function peaked in June (maximum) and
December (minimum). Such seasonal variation should provide an
annual modulation in the WIMP interaction rates and in the deposited
energy as a clear signature of the WIMP. However, the predicted
modulation is only a few percent of the unmodulated, average signal. To
reveal this rate modulation, large detector masses/exposures are
needed.

Such type of seasonal modulation in the spectra has been reported in
the DAMA experiment and attributed by the Collaboration to a WIMP
signal (see P.L. Belli's contribution to these Proceedings).

Preceding the DAMA experiment, there have been various attempts to
search for annual modulation of WIMP signals, starting as early as in
1991. COSME-1 and NaI-32 (Canfranc) (with 2 years statistics),
DAMA-Xe (Gran Sasso), ELEGANTS-V (Kamioka), and more recently
DEMOS (Sierra Grande) (with 3 years of statistics), are examples of
seasonal modulation searches with null results. However, these
experiments produced results which improved the $\sigma$(m) exclusion plots
and settled the conditions and parameters for new, more sensitive
searches.

\begin{figure}[htb]
\includegraphics[width=7.2cm]{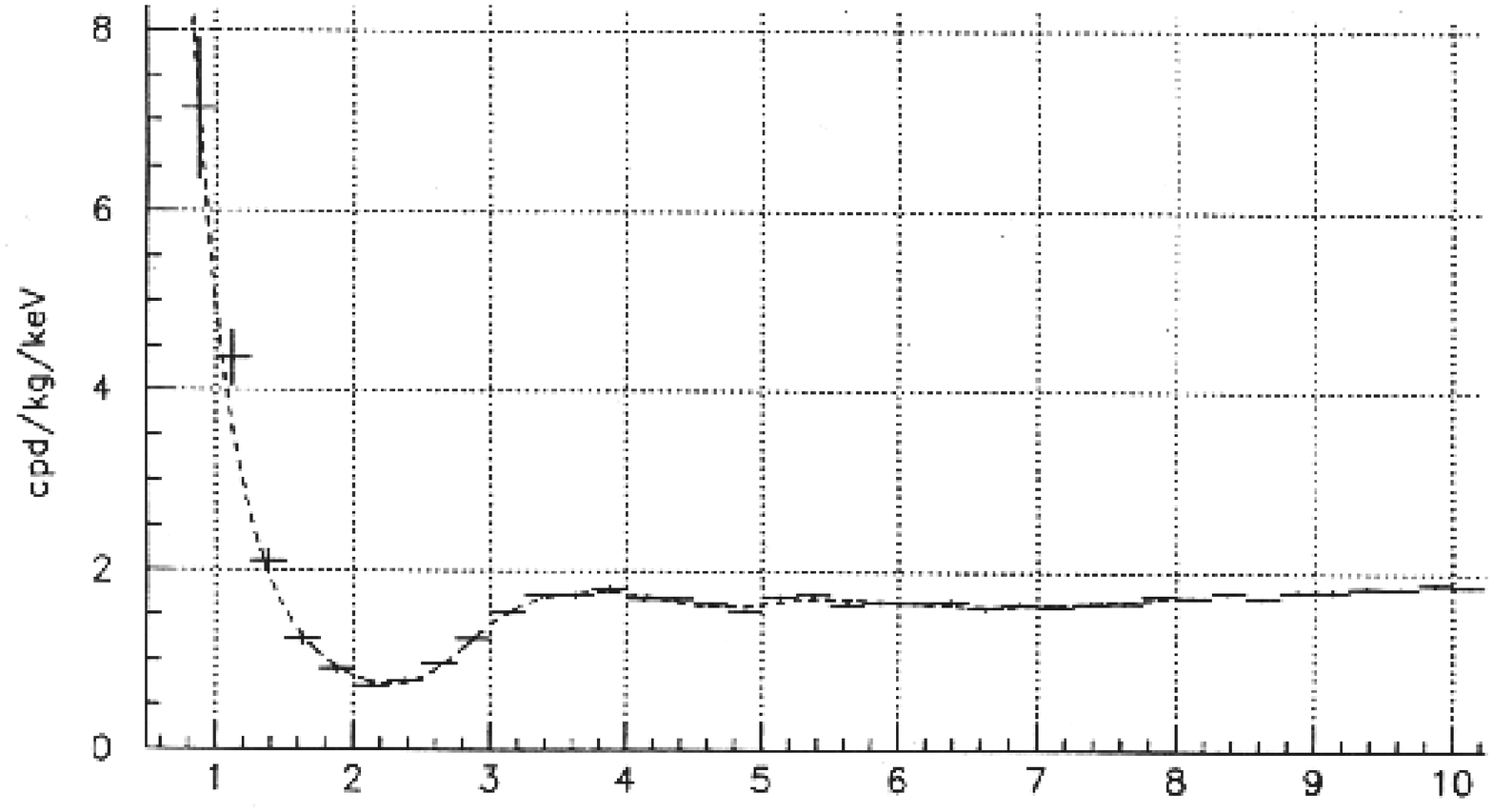}
\caption{Energy spectrum recorded with the multidetector set-up
of the DAMA experiment (single hit events)}
\label{fig:fig7}
\end{figure}

The DAMA experiment uses a set of 9 radiopure NaI crystals of 9.7 kg
each, viewed by two low-background PMT at each side through light
guides (10 cm long), coupled to the crystal. Special care has been taken
in controlling the stability of the main experimental parameters. A noise
removal is done by using the timing behaviour of noise and true NaI
pulses (with 40\% efficiency at low energy). The background after noise
removal is, averaging over the detectors: B (2--3 keV) $\sim$ 1--0.5
c/keV kg day and B (3--20 keV) $\sim$ 2 c/keV kg day. Notice the drop
in the first two channels, precisely where the expected signal is more
significant and, consequently, essential for deriving the reported
modulation effect. (The Pulse Shape Analysis is not used in the annual
modulation search). The multidetector energy spectrum of single hit
events (each detector has twelve detectors as active veto) has been
reported to this Workshop (Fig. \ref{fig:fig7}) (see P.L. Belli's
contribution to these Proceedings).

\begin{figure}[htb]
\includegraphics[width=7.2cm]{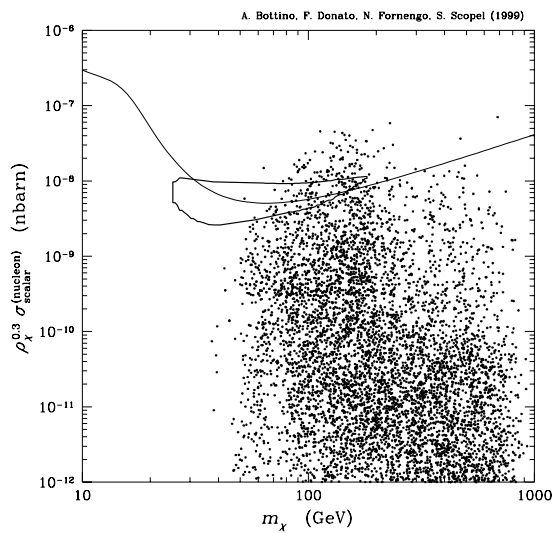}
\caption{Region ($\sigma$, m) of the positive modulation
effect found by DAMA.}
\label{fig:fig8}
\end{figure}

\begin{figure}[htb]
\includegraphics[width=7.2cm]{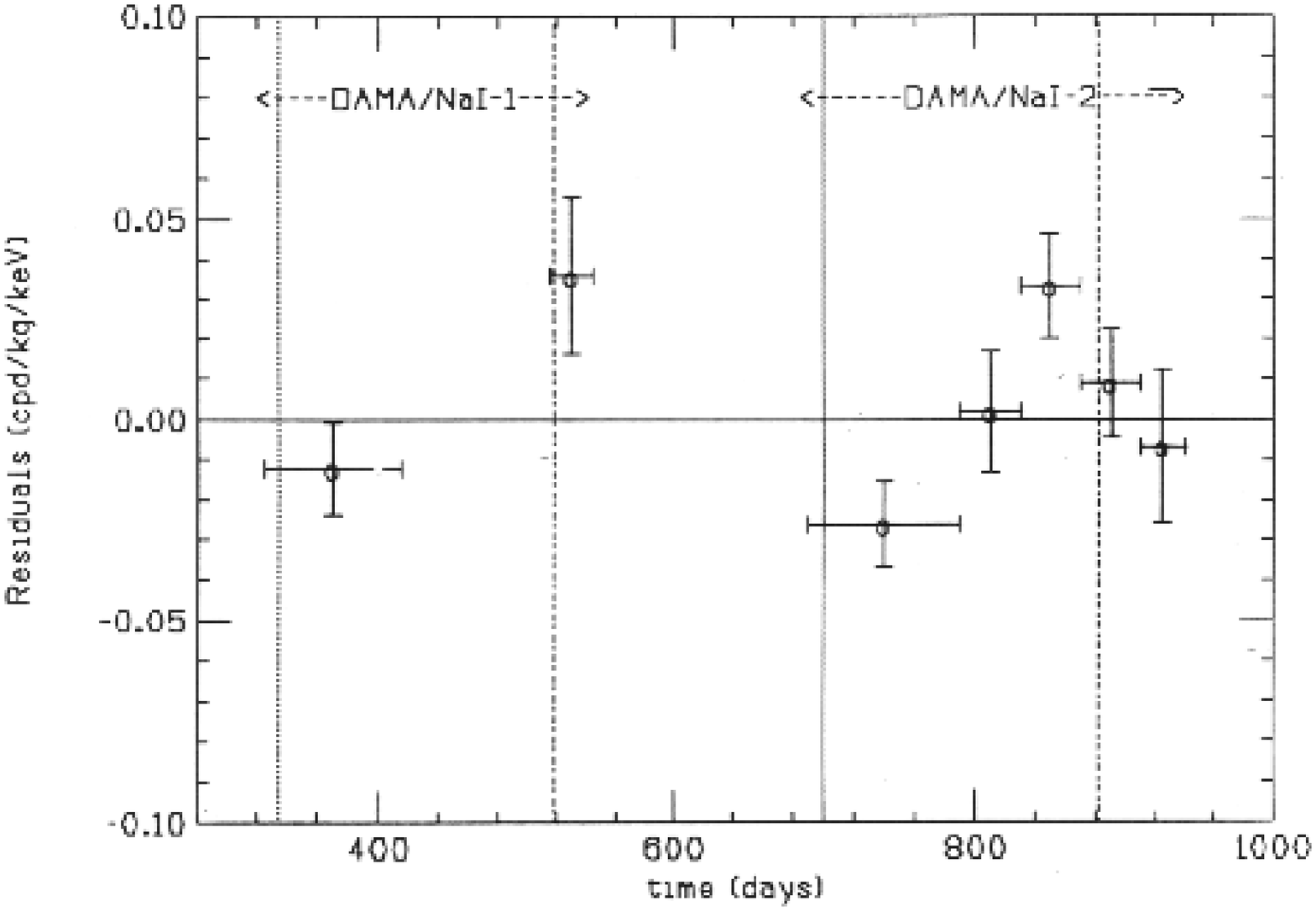}
\caption{Time evolution of DAMA signal}
\label{fig:fig9}
\end{figure}

At the time of TAUP 99, DAMA had issued the results of two runnings.
Run 1 reported at TAUP 97, extended over 1 month in winter and 2
weeks in summer, i.e. a total of 4549 kg day. Run 2, which used a
slightly different setup, extended from November to July, (one detector
for 90 days and eight detectors for 180 days) i.e. a total of 14962 kg
day. Both DAMA 1 and 2 results are compatible and consistent with
each other. A likelihood method applied to the total statistics of 19511
kg day provides a minimum for:
$m = \left( 59_{-14}^{+17} \right)$ GeV,
$\sigma^{\rm p} = \left( 7.0_{-14}^{+17} \right) \times 10^{-6}$ pb as the most
likely values of the mass and interaction cross-section on proton of an
``annual oscillating'' WIMP (at 99.6\% C.L.). Other statistical
approaches essentially agree with the likelihood result. Fig. \ref{fig:fig8}
shows the so-called DAMA region of the positive modulation effect and the
scattered plot of the MSSM prediction of $\sigma^{\rm p}$ in function
of m. The time evolution of the DAMA signal has been recently issued by
the collaboration indicating an oscillating trend in spite of the small
exposure time and the discontinuity of the two runnings (Fig. \ref{fig:fig9}).
The DAMA results have aroused great interest but also critical
comments related to various aspects of the experiment. One of the most
frequently heard has to do with the drop of background in the first
energy bins or to the peculiar look of the residual background after
subtracting the likelihood-emerging signal. Obviously the delivery of
new data is eagerly awaited.

Independently of other considerations and beyond any controversies
whatsoever, it is imperative to confirm the DAMA results by other
independent experiments with NaI (like ANAIS or ELEGANTS) and with
other nuclear target. The ZARAGOZA group is
preparing ANAIS (Annual Modulation with NaIs), consisting of 10 NaI
scintillators of 10.7 kg each, stored for more than ten years in
Canfranc, and recently upgraded for DM searches. It will be placed in
Canfranc within a shielding of electroformed copper and a large box in
Roman lead, plus a neutron screen and an active veto. The tests of a
smaller set are underway. Expected performances are an energy
threshold of 2 keV and a background at threshold of $\sim$ 2--3
c/(keV kg day).

The OSAKA group is performing a search with the ELEGANTS V NaI
detector in the new underground facility of Otho, with a huge mass of
NaI scintillators (760 kg) upgraded from previous experiments. A search
for annual modulation with null result has been presented to this
Workshop (see S. Yoshida's contribution to these Proceedings).

The DAMA $\sigma$(m) region should also be explored by the standard
method of comparing theory with the total time-integrated experimental spectrum
(without enquiring about possible variations in time). In fact, various
experiments are reaching the DAMA region (below
$\sim \sigma^{\rm p} \sim 10^{-5} - 10^{-6}$ pb for WIMPs of 40--80
GeV) which is itself half-excluded by a previous DAMA-0 running data
using PSA discrimination. For instance, CDMS has reached the upper
left corner and exclude it (see Fig. \ref{fig:fig2}  and R. Gaitskell's
contribution to these Proceedings), whereas the IGEX and H/M germanium
experiments are very close to it (with direct, non-stripped data).

Due to what is at stake, it is important to know what the prospects of
WIMP detection are trough the annual modulation signature for
planning the right experiments. In fact, to find an unambiguous,
reproducible and statistically significant modulation effect is, by now,
the best identifying label of a WIMP. Sensitivity plots for modulation
searches (presented to this Workshop) give the MT exposure needed to
explore $\sigma$,m regions using the annual modulation signature or,
equivalently, needed to detect an effect (should it exist) at a given C.L.,
due to a WIMP with a mass m and cross-section $\sigma^{\rm p}$.
Examples of sensitivity plots for Ge, NaI and TeO$_2$---and the
ensuing capability to explore $\sigma^{\rm p}$,m regions, are given to
illustrate the modulation research potential of some detectors (see S.
Scopel's contribution to these Proceedings) running or in preparation.

\section{Cryogenic Particle Detectors}
In the WIMP scattering on matter, only a small fraction of the energy
delivered by the WIMP goes to ionization, the main part being released
as heat. Consequently, thermal detectors should be suitable devices for
dark matter searches with quenching factors of about unity and low
effective energy threshold ($\rm E_{\rm vis} \sim \rm T$) as WIMPs
searches require. Moreover, bolometers which also collect charge (or
light) can simultaneously measure the phonon and ionization
(or scintillation) components of the energy deposition providing
a unique tool of background subtraction and particle identification.

There exist five experiments searching for direct interactions of WIMPs
with nuclei based on thermal detection currently running (MIBETA,
CDMS, EDELWEISS, CRESST, and ROSEBUD) another one,
CUORICINO, being mounted and a big project, CUORE, in preparation.

The CRESST (Cryogenic Rare Event Search with Superconducting
Thermometers) (MPI Munich / TUM Garching / Oxford, Gran
Sasso) detectors are four sapphire crystals (Al$_2$O$_3$) of
262 g each with a tungsten superconducting transition edge sensor.
The energy resolution and threshold obtained with an x-ray
fluorescence source are respectively 133 eV (at 1.5 keV) and 500 eV.
The background obtained in the Gran Sasso running is of about 10--15
counts/(keV kg day) above 30 keV, going down to 1 c/(keV kg day)
above 100 keV, whereas below 30 keV the spectrum is largely
dominated by noise and other spurious sources preventing to derive
exclusion plots.

Recently the collaboration has performed simultaneous measurements
of scintillation and heat, with a 6 g CaWO$_4$ crystal as absorber. The
preliminary results indicate a rejection of electron recoil events with an
efficiency greater than 99.7\% for nuclear recoil energies above 15
keV. Short term prospects for CRESST are the implementation of the
scintillation-phonon discrimination of nuclear recoils in a CaWO$_4$
detector of 1 kg (see J. Jochum's contribution to these Proceedings).

ROSEBUD (Rare Objects Search with Bolometers Underground)
[University of Zaragoza and IAS (Orsay)] is another sapphire bolometer
experiment to explore the low energy (300 eV--10 keV) nuclear recoils
produced by low mass WIMPs. It is currently running in Canfranc (at
2450 m.w.e.). It consists of two 25g and one 50g selected sapphire
bolometers (with NTD (Ge) thermistors) operating inside a small dilution
refrigerator at 20 mK. One of the 25g sapphire crystals is part of a
composite bolometer (2 g of LiF enriched at 96\% in $^6$Li glued to it)
to monitor the neutron background of the laboratory. The inner (cold)
shielding and the external one are made of archaeological lead of very
low contamination. The experimental setup is installed within a Faraday
cage and an acoustic isolation cabin, supported by an antivibration
platform. Power supply inside the cabin is provided by batteries and
data transmission from the cabin through convenient filters is based on
optical fibers. Infrared (IR) pulses are periodically sent to the bolometers
through optical fibers in order to monitor the stability of the experiment.
Pumps have vibration-decoupled connections.

\begin{figure}[htb]
\includegraphics[width=7.2cm]{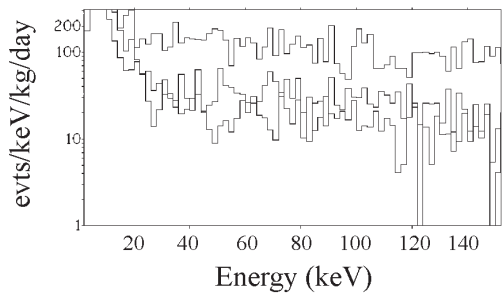}
\caption{ROSEBUD background spectrum}
\label{fig:fig10}
\end{figure}

The first tests in Canfranc have shown that microphonic and electronic
noise level is quite good, about 2nV/Hz$^{1/2}$ below 50Hz. The
bolometers were tested previously in Paris (IAS) showing a threshold of
300 eV and energy resolution of 120 (at 1.5 keV). Typical sensitivities
obtained (in Canfranc) are in the range of 0.3--1 $\mu$V/keV. Overall
resolutions of 3.2 and 6.5 keV FWHM were typically obtained in
Canfranc with the 50 g and 25 g bolometers, respectively, at 122 keV.
Low energy background pulses corresponding to energies below 5 keV
are seen. In the test runnings, the background obtained was as large as
120 counts/keV/kg/day around 40--80 keV. After various
modifications in the cryostat components, the background level of the
50 g bolometer stands about 15 counts/keV/kg/day from 20 to 80
keV. This progressive reduction is illustrated in  Fig. \ref{fig:fig10}.
Measurements of the radiopurity of individual
components continue with an ultralow background Ge at Canfranc and
their removal done when needed with the purpose of lowering the
background one more order of magnitude. The next step of the
ROSEBUD program will deal with bolometers of sapphire and
germanium, operating together to investigate the target dependence of
the WIMP rate (see P. de Marcillac's contribution to these Proceedings).

The Cryogenic Dark Matter Search Collaboration (CDMS)
(CfPA / UC Berkeley / LLNL / UCSB / Stanford / LBNL / Baksan / Santa
Clara / Case Western / Fermilab / San Francisco State) has developed
bolometers which collect also electron-holes carriers for discriminating
nuclear recoils from electron recoils. The electron-hole pairs are
efficiently collected in the bulk of the detector, but the trapping sites
near the detector surface produce a layer ($10-20 \mu \rm m$) of poor
charge collection, where surface electrons from outside suffer ionization
losses and fake nuclear recoils.

Two types of phonon readout have been developed. In the BLIP
(Berkeley Large Ionization and Phonon) detector, a NTD Ge thermistor
reads the thermal phonons in milliseconds. In the FLIP (Fast Large
Ionization and Phonon) detectors, non-equilibrium, athermal phonons
are detected (microsecond time scale) with superconducting transition
edge thermometers in tungsten. Current prototypes are BLIPs of 165 g
Germanium and FLIPs of 100 g Silicon. The FWHM energy resolutions
are 900 eV and 450 eV respectively in the phonon and charge channels
in BLIP detectors and about 1 keV in the FLIPs. The electron nuclear
recoil rejection in both detectors is larger than 99\% above 20 keV
recoil energy. Backgrounds below 0.1 c/(keV kg day) in the 10--20 keV
energy region have been obtained in recent runs.

\begin{figure}[htb]
\includegraphics[width=7.2cm]{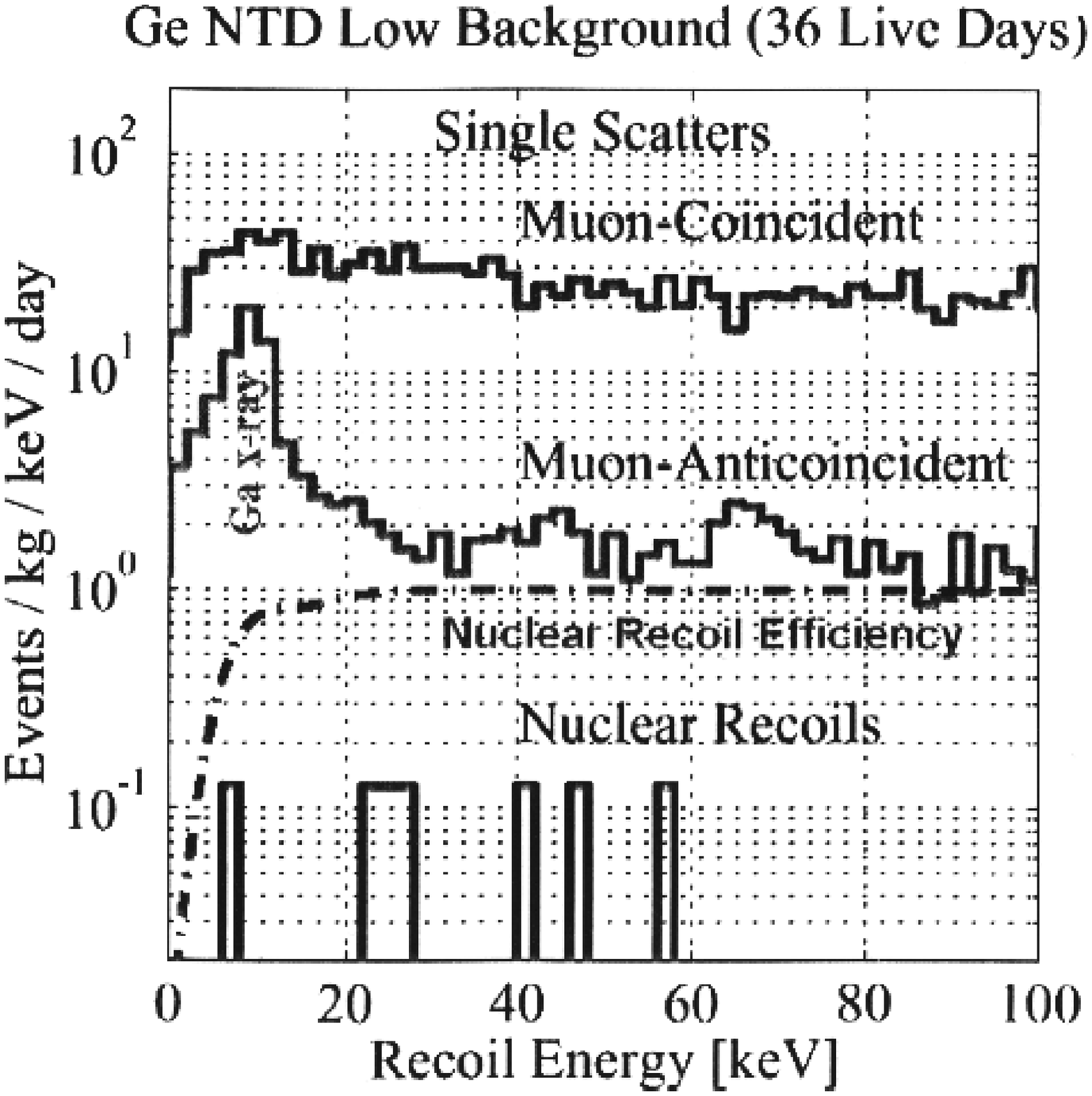}
\caption{CDMS low energy spectrum obtained in a recent run}
\label{fig:fig11}
\end{figure}

Following new developments achieved in the detectors, the surface
events have been successfully discriminated using their phonon rise
time: the low-charge collection events (surface electrons) have been
proved to have faster phonon rise time than the bulk events. A rise time
cut is applied to get rid of them. Results from a recent run are depicted
in Fig. \ref{fig:fig11}. In spite of the small masses and short time runs, CDMS
exclusion plots are competitive with much larger exposures of other
detectors. In fact, CDMS is now probing the DAMA region, as reported
to this Workshop (see Fig. \ref{fig:fig2} and R. Gaitskell's contribution to these
Proceedings). Projects of the CDMS Collaboration include the transfer of
the FLIP technology to germanium crystals of 250 g. Planned exposure
at Stanford (only 17 m of overburden) is of 100 kg day, with a
background goal of $\rm B=0.01$ c/keV kg day. The experiment will be
moved to Soudan along the year 2000 with twenty FLIP detectors of
Germanium (250 g each) and the background goal of $\rm B = \rm few
\times 10^{-4}$ c/keV kg day.

EDELWEISS (Orsay/Lyon/Saclay/LSM/IAP) has operated two 70 g HP
Ge bolometer in the Frejus tunnel with heat-ionization discrimination
getting similar results to that of the BLIP detectors of CDMS and so they
will not be repeated here. The background obtained is $\rm B = 0.6$
c/keV kg day in the $12-70$ keV region of the recoil energy spectrum.
The rejection is 98\% for surface events and $> 99.7$\% for internal
events. (For more details, see G. Chardin contribution to these
Proceedings).

The collaboration is preparing a small tower of three Ge bolometers of
70 g each, to be enlarged to other three of 320 g each. The background
goal is to get $\rm B = 10^{-2}$ c/keV kg day which seems to be at
hand. A second phase of EDELWEISS (2000--2001) will use a reverse
dilution refrigerator of 100 liters now under construction to host
$50-100$ detectors. Twenty Ge detectors of 300 g will be placed in the
next two years, expecting to improve the rejection up to 99.99\% and
get a background of $10^{-4}$ c/(keV kg day).

CUORE (Cryogenic Underground Observatory for Rare Events)
(Berkeley / Florence / LNGS / Leiden / Milan / Neuchatel/ South Carolina /
Zaragoza) is a project to construct a large mass (775 kg) modular
detector consisting of 1020 single bolometers of TeO$_2$ of dimensions
$5 \times 5 \times 5 \ \rm cm^3$ and 760 g each, with glued NTD Ge
thermistors, to be operated at 7 mK in the Gran Sasso Laboratory. A
tower of 14 planes consisting of 56 of those crystals with a total mass of
42 kg, the so-called CUORICINO detector, will be a first step in the
CUORE project.

Preliminary results of a 20 crystal array of tellurite bolometers (340 g
each) (MIBETA experiment) optimized for $2 \beta$ decay searches
show energy thresholds ranging from 2 to 8 keV (depending on the
detector) and background levels of a few counts per keV kg day in the
15--40 keV low energy region. CUORICINO is planned
as an extension of the MIBETA setup featuring more and larger crystals.
The sum of the 20 contemporary calibration spectra with a single
$^{232}$Th source shows that the array is indeed acting as a single
detector.

Four bolometers of the future CUORICINO array have been recently
tested in Gran Sasso. The results on the energy resolution in the
region of neutrinoless double beta decay of $^{130}$Te (2500 keV)
are about $5 \sim 8$ keV (see M. Pavan's contribution to these
Proceedings). Other values obtained are $\sim 2$ keV at 46 keV
and 4.2 keV at 5400 keV. Energy resolutions of 1-2 keV
and backgrounds of $10^{-2}$ c/(keV kg day) in the few
keV region can be expected.

\begin{figure}[t]
\includegraphics[width=7.2cm]{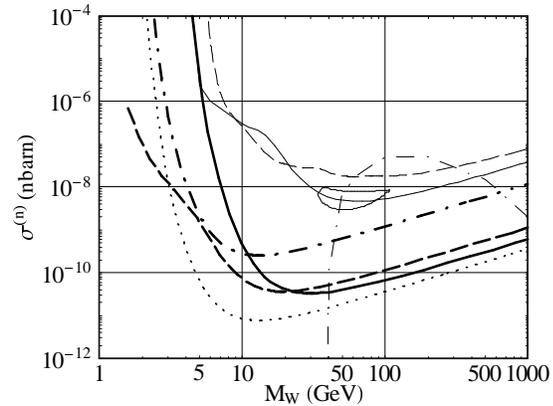}
\caption{Projection of exclusion plots expected for CUORICINO
(thick dashed), GEDEON (thick solid), CDMS Stanford (thick dot-dashed)
and CDMS Soudan (dots). Ge-combined (thin dashed) and DAMA (thin solid)
results are also shown as well as the MSSM region (thin dot-dashed)}
\label{fig:fig12}
\end{figure}

Fig. \ref{fig:fig12} shows the exclusion contour obtained from
running experiments (Ge, NaI), and the projections for
GEDEON, CUORICINO and CDMS assuming the parameter values expected in such
experiments.

\section{Where we stand and where we go}
 Unrevealing the nature of the dark matter is of uttermost importance in
 Cosmology, Astrophysics and Particle Physics. It has triggered a large
 experimental activity in searching for all its possible forms, either
 conventional or exotic. In particular, there exist various large
 microlensing surveys looking for dark baryons (EROS, MACHO,
 OGLE...) and a variety of observations searching for the baryonic
 component of dark matter. (See J. Us\'{o}n's contribution to these
 Proceedings).

As far as the exotic, non-baryonic objects are concerned, a few
experiments are looking (or project to look) for axions (RBF/UF,
LIVERMORE, KYOTO, CRYSTALS, CERN Solar Axion Telescope
Antenna...), as reviewed in P. Sikivie's contribution to these
Proceedings.

In the WIMP sector (see L. Roszkowski's contribution to these
Proceedings) to which this experimental overview is dedicated, there are
various large underground detector experiments (MACRO, BAKSAN,
SOUDAN, SUPER-K...) and deep underwater (ice) neutrino telescopes
(AMANDA, BAYKAL, ANTARES, NESTOR...) looking for (or planning to
look for) neutrino signals originated by the annihilation of WIMPs, as
well as some balloons and satellite experiments looking for antimatter of
WIMP origin, most of them included in these Proceedings. About thirty
experiments either running or being prepared are looking for WIMPs by
the direct way (COSME, IGEX, HEIDELBERG/MOS COW, ELEGANTS-V
and VI, DAMA, SACLAY, UKDMC, ANAIS, TWO-PHASE Xe, LqXe,
CASPAR, SIMPLE, MICA, DRIFT, CRESST, ROSEBUD, MIBETA,
CUORICINO, CDMS, EDELWEISS, ORPHEUS...), with conventional as
well as with cryogenic techniques... and some large projects with 100 to
1000 detectors (CUORE, GEDEON, GENIUS...) are being initiated. Their
current achievements and the projections of some of them have been
shown in terms of exclusion plots $\sigma^{\rm p}$ (m), which illustrate
the potential to investigate the possible existence of WIMP dark matter
in regions pretty close to where the supersymmetric candidates must appear.

After witnessing the large activity and progress reported to this
Workshop, it is clear that the main strategies recommended to search
for WIMPs have proved to be quite efficient to reduce the window of the
possible particle dark matter and to approach the zone of the more
appealing candidates and couplings. Examples of the achievements in
radiopurity, background identification or rejection, in low (effective)
threshold energy and efficiency, as well as in investigating the genuine
signatures of WIMPs, like modulation and directionality, have been
largely reported to TAUP 99 and reviewed selectively in this paper. The
conclusion is that these strategies are well focused and should be
further pursued. Finally, an annual modulation effect---supposedly
produced by a WIMP---is there, alive since the last TAUP 97, waiting to
be confirmed by independent experiments.

\section*{Acknowledgements}
I wish to thank the spokespersons of the experiments presented at
TAUP 99 for making available to me in advance their contributions as
well as other useful information about the status and plans of their
experiments. I am indebted to my collaborators I.G. Irastorza and S.
Scopel for their contribution to the making of the exclusion plots. The
kindness of my collaborators of COSME, IGEX and ROSEBUD for
allowing me to use the data from these experiments is warmly
acknowledged. Thanks are due also to my CUORICINO colleagues for
their permission to use internal information on the status and
preliminary results of the experiment. Finally, I thank Mercedes
Fat\'{a}s for her patience and skill in the composition of the text. The
financial support of CICYT (Spain) under grant AEN99-1033 and the
European Commission (DGXII) under contract ERB-FMRX-CT-98-0167
is duly acknowledged.

\end{document}